\documentclass[letterpaper]{article} % DO NOT CHANGE THIS
\usepackage{aaai25}  % DO NOT CHANGE THIS
\usepackage{times}  % DO NOT CHANGE THIS
\usepackage{helvet}  % DO NOT CHANGE THIS
\usepackage{courier}  % DO NOT CHANGE THIS
\usepackage[hyphens]{url}  % DO NOT CHANGE THIS
\usepackage{graphicx} % DO NOT CHANGE THIS
\usepackage{natbib}  % DO NOT CHANGE THIS AND DO NOT ADD ANY OPTIONS TO IT
\usepackage{caption} % DO NOT CHANGE THIS AND DO NOT ADD ANY OPTIONS TO IT
\usepackage{booktabs} % added by Yuzhe for Table format
\usepackage{makecell} % added by Yuzhe for Table format
\usepackage{amssymb} % added by Yuzhe for Table format
\usepackage{algorithm}
\usepackage{algorithmic}

\usepackage{xcolor}
\newcommand{\Mozhgan}[1]{\textcolor{black}{#1}}
\newcommand{\Romina}[1]{\textcolor{black}{#1}}
\newcommand{\Yueqian}[1]{\textcolor{black}{#1}}
\newcommand{\Yuzhe}[1]{\textcolor{black}{#1}}

\usepackage{newfloat}
\usepackage{listings}
\DeclareCaptionStyle{ruled}{labelfont=normalfont,labelsep=colon,strut=off} % DO NOT CHANGE THIS
\floatstyle{ruled}
\newfloat{listing}{tb}{lst}{}
\floatname{listing}{Listing}
\title{GenAI at the Edge: Comprehensive Survey on Empowering Edge Devices}

\author {
    Mozhgan Navardi\textsuperscript{\rm 1},
    Romina Aalishah \textsuperscript{\rm 1},
    Yuzhe Fu \textsuperscript{\rm 2},
    Yueqian Lin \textsuperscript{\rm 2},
    Hai Li \textsuperscript{\rm 2},
    Yiran Chen \textsuperscript{\rm 2},
    Tinoosh Mohsenin \textsuperscript{\rm 1}
}
\affiliations {
    \textsuperscript{\rm 1} Johns Hopkins University,  
    \textsuperscript{\rm 2} Duke University \\
    mnavard1@jhu.edu, raalish1@jhu.edu, yuzhe.fu@duke.edu, yueqian.lin@duke.edu, \\ hai.li@duke.edu, yiran.chen@duke.edu, Tinoosh@jhu.edu
}

\begin{document}

\maketitle

\begin{abstract}

\Mozhgan{Generative Artificial Intelligence~(GenAI) applies models and algorithms such as Large Language Model~(LLM) and Foundation Model (FM) to generate new data. GenAI, as a promising approach, enables advanced capabilities in various applications, including text generation and image processing. In current practice, GenAI algorithms run mainly on the cloud server, leading to high latency and raising security concerns. Consequently, these challenges encourage the deployment of GenAI algorithms directly on edge devices. However, the large size of such models and their significant computational resource requirements pose obstacles when deploying them in resource-constrained systems. This survey provides a comprehensive overview of recent proposed techniques that optimize GenAI for efficient deployment on resource-constrained edge devices. For this aim, this work highlights three main categories for bringing GenAI to the edge: software optimization, hardware optimization, and frameworks. The main takeaways for readers of this survey will be a clear roadmap to design, implement, and refine GenAI systems for real-world implementation on edge devices.}

\end{abstract}

\section{Introduction}

Generative Artificial Intelligence (GenAI) has become a promising solution in text generation, image synthesis, and multimodal content creation. These developments often rely on large-scale models such as Large Language Models (LLMs) %and Foundation Models (FMs) 
that achieve remarkable performance but demand large computational and memory resources. Traditionally, these models run on powerful cloud servers, which introduces latency, dependency on network connectivity, and potential privacy risks. As real-time applications and data security become ever more critical, there is a growing push to embed GenAI functionalities directly into edge devices~\cite{nezami2024generative, mozhgan2024metatinymlv2}.

However, implementing high-intensive models on the edge presents significant challenges~\cite{pourmehrani2024fat, uttej2024resourcev2, humes2023squeezedv2}. 
Edge devices, including drones~\cite{iscas2023-mozhganv2}, and autonomous systems~\cite{reprohrl2023-tejaswiniv2} benefit significantly from the GenAI capabilities on devices. For instance, 
drones can generate real-time terrain analysis in remote areas, 
Autonomous systems can enhance decision-making through local models. Wearable health monitoring could generate personalized insights from biometric data while ensuring privacy through local data processing. To support these applications, specialized edge hardware such as NVIDIA Jetson, and Qualcomm AI Engine have been developed to handle the computational demands of GenAI while maintaining efficiency.

This situation calls for innovative approaches in software optimization including model compression, Neural Architecture Search (NAS). 
In parallel, hardware optimization including specialized accelerators, attention optimization, and dedicated frameworks address computational and energy constraints at the edge~\cite{asmer2024energyv2}. These strategies not only reduce model size and inference latency but also address privacy concerns when deploying complex models on edge devices~\cite{mozhgan2024metatinymlv2}. This paper aims to survey existing methods and provide extensive details on implemented GenAI techniques on edge devices. To the best of our knowledge, there is no dedicated survey on GenAI at the edge. By reviewing state-of-the-art techniques from top-tier conferences and journals, this work offers a  roadmap for researchers seeking to apply GenAI in edge.
The main category of the paper is organized as follows:

\begin{itemize}
    \item \textbf{Software Optimization:}
    Discusses key strategies for adapting GenAI models to edge devices, including model compression methods (pruning, quantization, and knowledge distillation), NAS, and open-source GenAI models. %and Federated Learning (FL).

    \item \textbf{Hardware Optimization:}
    Explores %specialized 
    hardware accelerators and attention optimization to highlight how they meet GenAI’s computational demands while addressing power and resource constraints on edge devices.

    \item \textbf{Frameworks:}
    Reviews frameworks %like TensorRT 
    to improve inference latency, memory, and overall energy efficiency.

\end{itemize}

\begin{figure*}
    \centering
    \includegraphics[width=0.95\textwidth]{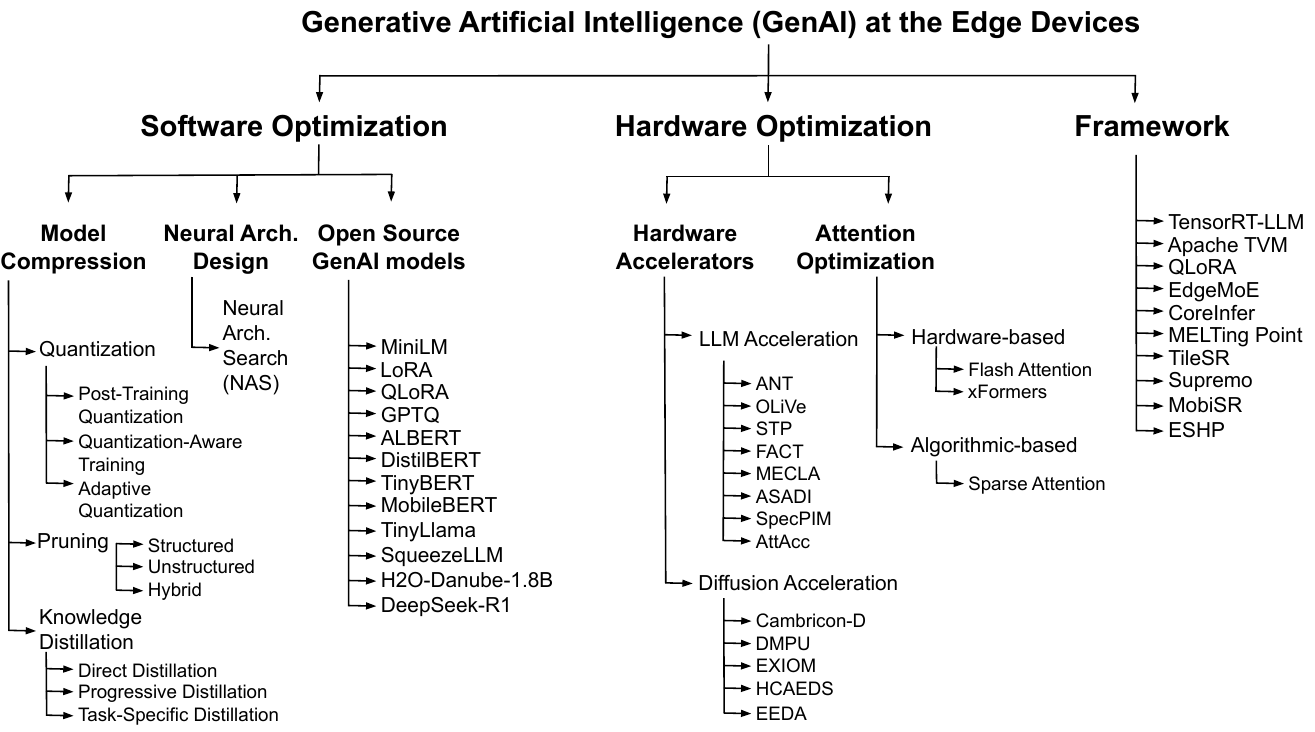}
    %\vspace{-0.8cm}
    \caption{Illustration of the flow of GenAI at the edge}
    \vspace{-0.5cm}
\end{figure*}

\section{Software Optimization}
\label{sec:1}

\subsection{Model Compression}
\Yueqian{The rapid advancement of GenAI models, while ushering in unprecedented capabilities, has also given rise to increasingly large model architectures that present significant deployment challenges \cite{guo2024survey}.  
Early attempts to address these challenges explored distributed mobile computing systems that could partition model computation across multiple devices \cite{mao1,mao2}.}

\Yueqian{This challenge has since prompted extensive research in model compression techniques, which have evolved along three principal directions to enable broader deployment and accessibility. Firstly, quantization techniques have achieved remarkable efficiency through reduced precision representations, particularly through enhanced activation distribution handling and hardware-optimized strategies. Secondly, methodologies for pruning have advanced from rudimentary magnitude-based techniques to sophisticated hardware-aware structured approaches, enabling considerable model reduction while preserving architectural integrity.  Thirdly, knowledge distillation has evolved to incorporate progressive frameworks and multi-teacher architectures, showing particular promise in task-specific applications. Contemporary research emphasizes hardware-aware compression strategies and architecture-specific solutions. While these advancements have enabled the deployment of foundation models with competitive performance metrics, the fundamental challenge persists in optimizing the compression-performance trade-off for edge deployment scenarios.}

\textbf{Quantization}
\Yueqian{Model quantization has emerged as a critical technique for deploying large-scale GenAI models on resource-constrained edge devices. Quantization approaches are broadly categorized into post-training quantization (PTQ) and quantization-aware training (QAT). PTQ methods like OPTQ \cite{frantar2023optq} and AWQ \cite{lin2023awq} directly convert trained model parameters to lower precision formats, while QAT approaches such as EdgeQAT \cite{shen2024edgeqat} incorporate quantization effects during training. PTQ methods are generally preferred due to their computational efficiency, though recent advances in both approaches have enabled effective compression through sophisticated handling of weight and activation distributions. When applied to LLMs, unique challenges emerge from their heavy-tailed weight distribution. Methods like SmoothQuant \cite{xiao2023smoothquant} and OliVe \cite{oliVe} address this through distribution smoothing and outlier handling techniques. Mixed-precision approaches \cite{chen2024channel} have shown promise by automatically determining optimal bit widths for different model components based on their quantization sensitivity. Recent work like OneBit \cite{xu2024onebit} and BitNet \cite{wang2023bitnet} has pushed boundaries by demonstrating viable 1-bit quantization through sophisticated distribution-aware schemes. However, significant challenges remain in maintaining generation quality under extreme compression and developing efficient training methods for quantized LLMs on edge devices \cite{egiazarian2024extreme}.}

\Yueqian{Diffusion models present their own set of quantization challenges, particularly in handling varying activation distributions across diffusion steps. Approaches like Q-DM \cite{q-dm}, PTQD \cite{ptqd}, and Q-Diffusion \cite{q-diffusion} tackle the challenge of varying activation distributions across diffusion steps through adaptive calibration and noise-aware quantization. Specialized temporal-aware quantization methods \cite{NEURIPS2023_983591c3,Huang_2024_CVPR} have been developed to handle the unique challenges of the iterative denoising process. Current research focuses on effectively handling dynamic activation ranges and balancing compression ratios with generation quality for edge deployment of diffusion models \cite{yao2024timestep}.}

\textbf{Pruning}
\Yueqian{Model pruning methods can be broadly categorized into structured and unstructured approaches, each with distinct trade-offs between compression efficiency and hardware compatibility. These techniques have shown particular promise in compressing large-scale generative models while maintaining performance for edge deployment. The field of LLM pruning has recently witnessed several novel approaches. Structured pruning methods like LLM-Pruner ~\cite{ma2023llmpruner} and edge-optimized approaches ~\cite{khiabani2025optimizing} achieve 2$\times$ speedup with minimal performance degradation by removing entire structural components.% such as attention heads. 
Unstructured approaches like SparseGPT \cite{frantar-sparsegpt} enable up to 60\% sparsity in large-scale models, while recent advances in modality-specific pruning techniques have shown promising results across speech, vision, and multimodal domains, with methods like SpeechPrune \cite{lin2024speechprune} achieving up to 80\% pruning rates while maintaining %task 
performance. Hardware-aware methods have become increasingly crucial, as exemplified by Flash-LLM \cite{xia2023flashllm}, which achieves 3$\times$  inference speedup through unstructured sparsity-aware system optimization. Semi-structured pruning methods such as E-Sparse \cite{li2023sparse} further advance this direction by leveraging N:M sparsity patterns to maintain hardware compatibility while achieving high compression rates on edge devices.}

\Yueqian{In the context of diffusion models, methods like Diff-Pruning \cite{fang2023structural} achieve approximately 50\% reduction in FLOPs by leveraging Taylor expansion over pruned timesteps while maintaining generative quality. Specialized approaches like LD-Pruner \cite{Castells_2024_CVPR} implement task-agnostic pruning strategies for Latent Diffusion Models, while DiP-GO \cite{zhu2024dipgodiffusionprunerfewstep} demonstrates 4.4$\times$  speedup on Stable Diffusion without requiring retraining. Recent work combines gradient-based pruning for mask matrix continuity \cite{wan2025pruningsparsediffusionmodels} with strategic data pruning \cite{briq2024datapruninggenerativediffusion}, showing particular promise for edge deployment where both computational efficiency and generation quality are critical \cite{yan2024hybrid}.}

\textbf{Knowledge Distillation.}
\Yueqian{Knowledge Distillation~(KD) has emerged as a crucial paradigm for deploying GenAI models 
on edge devices, with distinct approaches developed for different model architectures to balance model capabilities with computational constraints. The application of KD to language models has led %given rise 
to a variety of approaches. These can be categorized into white-box and black-box methods. White-box KD enables student models to match both final predictions and internal representations when the teacher model is open-source (e.g., LLaMA \cite{touvron2023llamaopenefficientfoundation}), while black-box KD works with closed-source models (e.g., GPT-4 \cite{openai2024gpt4technicalreport}) through API calls \cite{liu-etal-2024-evolving}. Notable advances include MiniLLM \cite{gu2024minillm}, which introduces a reversed Kullback-Leibler divergence objective to stabilize student updates, and instruction-following distillation approaches that have produced efficient open-source models like Vicuna \cite{vicuna2023} and Koala \cite{koala_blogpost_2023}. Recent work in instruction-following KD %distillation 
has enabled compact yet capable models through supervised fine-tuning \cite{wu-etal-2024-lamini}, while advanced applications like RLAI feedback \cite{lee2023rlaif} demonstrate the potential for model alignment through distillation. Adaptive distillation methods have further enhanced this field 
by dynamically adjusting the distillation process based on input complexity, allowing student models to focus learning where improvement is most 
needed~\cite{liang2024dynamic}.}

\Yueqian{In the domain of diffusion models, KD primarily focuses on accelerating sampling speed to address the challenge of high inference latency. Progressive distillation \cite{salimans2022progressive} represents an %breakthrough 
approach that iteratively halves sampling steps (e.g., from 1000 to 1), enabling efficient edge deployment while maintaining generation quality. Single-step approaches \cite{luhman2021knowledge} further compress diffusion teachers into one-step generators, although this requires careful balance between efficiency and generation fidelity. Teacher-free acceleration methods like DPM-Solver \cite{NEURIPS2022_260a14ac} and consistency models \cite{pmlr-v202-song23a} demonstrate effective inference cost reduction without extensive re-training. Recent advances include two-stage approaches \cite{meng2023distillation} for text-conditional models and score distillation sampling \cite{poole2023dreamfusion} for 3D generation, showcasing the versatility of distillation in different applications. Also, generative dataset distillation using models like SDXL-Turbo with class-specific prompts has achieved superior images per class ratios in recent benchmarks \cite{su2024diffusion}, offering new possibilities for efficient model training and deployment.}

\subsection{Neural Architecture Design}

Efficient neural architecture design has emerged as a critical research direction to address the increasing complexity and resource demands of modern models, particularly for edge devices \cite{howard2017mobilenets, NASsurvey2019}. 
By automating the generation of network architectures while considering specific hardware and constraints, computational overhead, required memory, and power consumption have been improved, while maintaining model performance.

\textbf{Neural Architecture Search (NAS).}~Neural Architecture Search (NAS)~\cite{NASfirst2016, NASsurvey2019} serves as a powerful framework to automate the design of optimal model topologies with strict latency, memory, or power budgets. By systematically exploring a predefined search space such as varying layer depth, width, or connection patterns. NAS algorithms can discover specialized architectures that outperform traditional solutions. In~\cite{NASfirst2016}, they have proposed the first NAS using reinforcement learning (RL) to determine optimal Recurrent Neural Network~(RNN) parameters. Subsequently, this idea was extended to Convonotional Neural Network~(CNNs) in \cite{NASextended2018}, where the authors integrated a Sequential Model-Based Optimization (SMBO) approach with a reinforcement mechanism for cell-based searches to find the best configuration.

In the context of GenAI, %Generative AI, 
where large models often dominate in tasks such as text generation or image synthesis, NAS-driven architectures present a promising route to achieve efficiency.
There are a
limited number of work on NAS in the field of transformers~\cite{liu2024mobilellm}.
FL-NAS~\cite{FLNAS2024} have proposed an approach which leverages LLM to find high-performance DNNs for resource-constrained systems. Moreover, work in~\cite{10646428} proposed a LLM-based methodology for NAS technique in Edge devices. 
Puzzle~\cite{bercovich2024puzzle} proposed an LLM optimized for inference using NAS under hardware constraints, achieving a 2.17x inference throughput speedup.

\subsection{Open-Source GenAI models}
%\textbf{Lightweight Transformers}
\Romina{The recent advancements in reasoning capabilities of models such as DeepSeek-R1 \cite{deepseek2025} emphasize the power of open research development.} %As Yann LeCun remarked:  \textit{“Open source models are surpassing proprietary ones.”}}
\Romina{DeepSeek-R1 \cite{deepseek2025} has profited significantly from open-source tools like PyTorch and Meta's Llama \cite{touvron2023llamaopenefficientfoundation}.} 
\Romina{One of the key contributions to the advancement in GenAI is open-source innovations, specifically for edge scenarios in which the resources are limited. In these cases, smaller model sizes and less latency besides not losing performance are the main considerations. Therefore, researchers explored various compression methods, leading to models like DistilBERT \cite{sanh2019distilbert}, TinyBERT \cite{jiao2019tinybert}, ALBERT \cite{lan2019albert},  MobileBERT \cite{sun2020mobilebert}, MiniLM \cite{wang2020minilmdeepselfattentiondistillation}, and MiniLMv2 \cite{wang2021minilmv2multiheadselfattentionrelation} each using techniques such as knowledge distillation, parameter sharing, or factorization to make large models smaller while maintaining strong performance.}

\Romina{Beyond these compression-based strategies that are already covered in the previous sections, novelties in architecture further improved efficiency. Reformer \cite{kitaev2020reformer} introduced locality-sensitive hashing for attention and reversible residual layers, enabling near-linear complexity for longer sequences. Meanwhile, GPT-NeoX-20B \cite{black2022gpt}, LLaMA \cite{touvron2023llamaopenefficientfoundation}, and LLaMA2 \cite{touvron2023llama2} showed how LLMs could be developed and released collaboratively, making it easier for edge-focused adaptations. %Making different versions of these models available, inspired some on efficiency as well as the feasibility of using by others. 
Even smaller-scale of these projects such as TinyLlama \cite{tinyllama2024} and H2O-Danube-1.8B \cite{h2odanube2024} now offer compact language models tailored to edge constraints, continuing the trend of collaborative research. Similarly, research on instruction tuning \cite{won2022finetuned}, which trains models to handle various tasks by exposing them to different instructions, reinforced the importance of building flexible and open-source foundations for further innovation.}

\Romina{Researchers have further built on open releases to develop conversational systems, including Alpaca \cite{alpaca2023}, Koala \cite{koala2023}, and Vicuna \cite{vicuna2023}, each developed by fine-tuning LLaMA \cite{touvron2023llamaopenefficientfoundation} on curated datasets, all demonstrating competitive performance against models like ChatGPT and Bard. These models have also served as benchmarks for edge-focused projects such as SqueezeLLM \cite{kim2024squeezellm}, which introduces a post-training quantization framework to compress LLMs for more efficient inference, focusing on reducing memory bandwidth, outperforming methods like GPTQ \cite{frantar2022gptq}, AWQ \cite{lin2023awq}, and SpQR\cite{dettmers2023spqr}. In parallel, techniques like LoRA (Low-Rank Adaptation) \cite{hulora2021} have reduced the cost of fine-tuning large models, accelerating domain-specific deployments. Later, QLoRA \cite{dettmers2023qlora} tried to fine-tune a large model on a single GPU by reducing memory usage by quantizing the quantization constants and using this technique.}
\Romina{Taken together, several open-source LLMs have been developed, and some of them are compressed to reduce their size and improve efficiency. 
These include MPT-7B \cite{MosaicML2023Introducing}, which implements a 7B-parameter architecture designed for commercial applications; DLite \cite{ai2023dlite}, which scales from 124M to 1.5B parameters; and RedPajama-INCITE \cite{together2023redpajama}, which spans 3B to 7B parameters.
Open-source models and innovations can be valuable for resource-constraint applications, and be fine-tuned for specific tasks to improve their performance.}

\section{Hardware Optimization}
\label{sec:2}
\subsection{Hardware Accelerators}

%\Yuzhe{ Yuzhe please add subsections and then start adding text to each subsection ...}

\begin{table*}
\centering
\normalsize
\caption{\Yuzhe{Hardware Accelerator for GenAI}}
\scalebox{0.8}{
\begin{tabular}{lcccccccc}
\toprule[1.5pt]
Accelerator & Year & Platform & Technology & Networks & Sparsity/Quantization & Peak Energy Efficiency (TOPS/W)\\
\midrule
EXION \cite{heo2025exion} & 2025 & ASIC simulator & 14nm & SD/DiT & \checkmark / \checkmark @INT12 &  $11.53$\\
\midrule
HCAEDS \cite{guo2024hcaeds} & 2024 & CIM tapeout & 28nm & SD & - / \checkmark @INT10/BF16 & $74.34$\\
\midrule
DMPU \cite{qin2024dmpu} & 2024 & ASIC tapeout & 22nm & DDPM & \checkmark / - & $52.01$\\
\midrule
EEDA \cite{yoo2024eeda} & 2024 & ASIC tapeout & 28nm & SD & - / \checkmark @HYP8 & $4.96$\\
\midrule
% SDA & 2024 & FPGA & - & SD & - / \checkmark @W4A8 & $6\times$ V100\\
% \midrule
Cambricon-D \cite{kong2024camb}& 2024 & ASIC simulator & 7nm & SD & \checkmark / \checkmark @INT3/FP16 & $13.34$\\
\midrule
AttAcc \cite{park2024attacc} & 2024 & CIM simulator & 7nm & LLaMA/GPT-3 & - / - & $2.67\times$ DGX A100\\
\midrule
SpecPIM \cite{li2024specpim} & 2024 & CIM simulator & - & LLaMA/OPT & - / - & $6.7\times$ A100\\
\midrule
ASADI \cite{li2024asadi}& 2024 & CIM simulator & 28nm & GPT-2/BERT & \checkmark / - & -\\
\midrule
MECLA \cite{qin2024mecla} & 2024 & ASIC simulator & 28nm & LLaMA/BERT & - / \checkmark @INT8 & $7.09$\\
\midrule
% FlightLLM & 2024 & FPGA & - & LLaMA/OPT & - / \checkmark @INT8  & $6.7\times$ V100\\
% \midrule
STP \cite{tambe2023STP} & 2023 & ASIC tapeout & 28nm & BERT & - / \checkmark @FP4 & $18.1$\\
\midrule
OliVe \cite{oliVe} & 2023 & ASIC simulator & 22nm & GPT-2/OPT/BERT & - / \checkmark @Adaptive 4bit & $4\times$ GOBO \cite{zad2020gobo}\\
\midrule
FACT \cite{qin2023FACT} & 2023 & ASIC simulator & 28nm & BERT & \checkmark / \checkmark @INT8  & $4.39$\\

\bottomrule[1.5pt]
\end{tabular}}
\label{tab:accForGenAI}
\end{table*}

\Yuzhe{Hardware accelerators are typically designed through the software and hardware co-design for specific networks. Algorithmically, data sparsity is enhanced by pruning, and model compression, such as quantization, reduces network size. On the hardware side, specific architectures are designed to bypass sparse or redundant computations, increase data reuse, and minimize data movement, thus enabling energy-efficient acceleration on edge devices. Generative AI (GenAI) includes GAN, LLM, and Diffusion models. While extensive hardware work has focused on optimizing GAN models \cite{chen2018regan, kim2021gan, kang2021ganpu}, recent trends have shifted toward LLM and Diffusion models, driving further hardware research in GenAI. This section reviews recent efforts in optimizing hardware accelerator for LLM and Diffusion networks, with representative works summarized in Table \ref{tab:accForGenAI}.}

\textbf{LLM Acceleration} \Yuzhe{LLM models have diverse distributions at the tensor or channel levels, numerous studies leverage customized data types to accommodate this challenge. For example, ANT~\cite{guo2022ant} introduces a novel data type and employs an adaptive mechanism to determine the most appropriate type for each tensor from a predefined set. Expanding on ANT, OliVe~\cite{oliVe} proposes an outlier-victim pair approach, which provides a more precise representation of outlier distributions in LLM models. Both ANT and OliVe incorporate specialized decoders and multiply-accumulate (MAC) units to optimize their arithmetic computation processes for LLMs. Some studies focus on reducing redundant computations in LLM models to improve the energy efficiency during inference. STP \cite{tambe2023STP} proposes a computation-skipping strategy and dynamic data path reconfiguration based on entropy, achieving high energy efficiency with minimal accuracy loss. Furthermore, it has been observed that linear projections contribute significantly to the memory footprint and latency in LLM models. FACT \cite{qin2023FACT} introduces an eager prediction method with a leading-one detector and log-based inner-product estimation, reducing computations in both attention and linear projections. MECLA \cite{qin2024mecla} surpasses FACT by decomposing large matrices into smaller sub-matrices to minimize off-chip memory access and re-associating data on-chip for better reuse.}

\Yuzhe{Recently, Computing-in-Memory (CIM) becomes a prominent approach for LLM acceleration. CIM accelerators offer significant energy efficiency gains, particularly for general matrix-matrix multiplication (GEMM) operations. Existing studies typically leverage CIM architectures to accelerate the attention mechanism, while relying on CPUs or GPUs to handle other operations. ASADI \cite{li2024asadi} introduces a sparse attention paradigm based on diagonal compression (DIA) format, enabling highly parallel computation on CIM processors. SpecPIM \cite{li2024specpim} accelerates speculative inference in LLM by optimizing resource allocation in CIM-enabled heterogeneous systems, while AttAcc \cite{park2024attacc} accelerates batched LLM inference on CIM/NPU heterogeneous systems. Given these developments, it is expected that CIM-based accelerators for LLM models will become more prevalent in the future.}

\textbf{Diffusion Acceleration} \Yuzhe{Diffusion networks have made significant progress recently in various GenAI tasks, with %significantly
different network architecture from LLM models. These networks generate images or videos through multiple iterations of denoising operations, with highly similar images in consecutive iterations. Consequently, hardware optimizations often leverage inter- and intra-iteration similarity to accelerate Diffusion networks, typically through differential computing and skipping redundant computations. }

\Yuzhe{Cambricon-D \cite{kong2024camb} introduces an approximate ReLU in the Stable Diffusion (SD) network, enabling differential computing for nonlinear functions and addressing the memory overhead associated with full-precision nonlinear calculations in traditional differential computing architectures. DMPU \cite{qin2024dmpu} observes that many pixels exhibit minimal changes between consecutive time steps in Diffusion models, and thus proposes a semantic-segment sparse convolution along with a trivial attention exponent inheritance method to skip redundant computations in both the convolution and attention mechanisms, significantly enhancing the energy efficiency. EXION \cite{heo2025exion} presents an FFN-Reuse algorithm that can be applied across iterations, along with an improved eager prediction method for predicting attention scores, which reduces redundant computations and boosts throughput. HCAEDS \cite{guo2024hcaeds} is the first heterogeneous CIM chip designed for Diffusion models, incorporating a Sign-Magnitude radix-8 Booth CIM macro for integer data and a four-operand exponent CIM macro for floating-point data, achieving a high energy efficiency.}

\Yuzhe{Numerous GenAI hardware studies \cite{kong2024camb, yoo2024eeda, yang2024sda, wang2024dtrans} have observed that nonlinear functions (such as softmax, GeLU, etc.) can introduce significant latency overhead during the hardware acceleration. These studies optimize nonlinear functions to enhance overall throughput. Additionally, some studies \cite{fu2024softact, dong2024NLO, ste2021softermax, yan2019cim_non} have focused specifically on optimizing nonlinear functions and have designed specialized hardware to facilitate network inference. All of these studies indicate a potential research trend on optimizing nonlinear functions in GenAI networks. Combined with techniques such as eliminating redundant computations and data compression, these approaches can enhance hardware acceleration and improve energy efficiency for GenAI systems.}

\subsection{Attention Optimization}

\Romina{Transformers have become the backbone of many GenAI models, but their multi-head self-attention mechanism can dominate runtime and memory usage. Therefore, researchers have explored a range of strategies to optimize attention on \emph{hardware} %(GPU)} 
and \emph{algorithmic} levels.}
%memory (

\textbf{Hardware-based.}~\Romina{%One example is 
FlashAttention \cite{dao2022flashattention} %, which 
reorders attention operations to reduce the number of reads and writes between GPU high bandwidth memory (HBM) and on-chip static RAM (SRAM) by splitting queries, keys, and values into smaller blocks, recomputing attention on-chip during the backward pass, and fusing multiple GPU kernels into one. Built on this, FlashAttention-2 \cite{dao2023flashattention2} takes the foundation of memory efficiency and adds better parallelism and work distribution to further increase speed and GPU utilization, especially for longer sequences. Then, FlashAttention-3 \cite{dao2024flashattention3} introduces asynchrony and low-precision computation to further optimize the attention mechanism for modern GPU architectures, which allows for even higher performance and efficiency, along with reduced error for low-precision (FP8) computing. Besides these, xFormers \cite{xformer}, a PyTorch-based library, provides a collection of optimized attention and Transformer blocks, including custom GPU kernels and memory-efficient attention implementations.}

\textbf{Algorithmic-based.}~\Romina{Work on sparse attention reduces the quadratic complexity of self-attention by ignoring parts of the input that do not affect the result significantly. 
Child et al. \cite{child2019sparse} pioneered this approach by limiting attention to strided patterns using sparse factorizations of the attention matrix to reduce computation cost while maintaining performance on sequence models. Subsequent techniques like Longformer \cite{beltagy2020longformer} by using a combination of sliding window local attention and task-motivated global attention, Big Bird \cite{zaheer2020bigbird} by combining random, windowed, and global attention to create a sparse attention mechanism, and Linformer \cite{wang2020linformer} by decomposing attention with linear projections to achieve linear complexity introduced various structured sparsity patterns. Meanwhile, Choromanski et al. \cite{choromanski2021performer} developed performer, which uses random feature maps to approximate the softmax function, reducing its time complexity from $\mathcal{O}(n^2)$ 
to $\mathcal{O}(n)$.}

\section{Frameworks}
\label{sec:3}

\Romina{Deploying GenAI models on edge devices might bring challenges because of limited computational power, memory, and latency requirements. To address these constraints, researchers have explored various techniques that simplify computations at both the graph and operator levels. By fusing kernels, reducing redundant operations or parameters, and customizing algorithms to the hardware, these methods enable fast inference for tasks such as large language modeling, super-resolution, and more.}

\Romina{%Solutions such as 
NVIDIA TensorRT and Apache TVM are pioneered compiler-based optimizations by combining graph-level fusion and quantization with lower latency. Likewise, Google’s EdgeTPU and Coral stacks enable rapid deployment of compressed models through low-power hardware and software stack. TensorRT-LLM \cite{tensorrtllm} is also a specialized toolkit for accelerating LLM inference on GPUs, including optimized CUDA kernels for attention computations, inflight batching, and quantization.}

\Romina{Beyond these compilers, researchers have developed frameworks customized for various GenAI workloads. For instance, Yi et al. proposed EdgeMoE \cite{yi2023edgemoe}, an engine specifically optimized for Mixture-of-Experts (MoE) language models. By using expert-wise bitwidth adaptation, it supports models with a large number of parameters on edge devices to reduce inference times substantially. Wang et al. introduced CoreInfer \cite{wang2024coreinferacceleratinglargelanguage}, achieving over 10$\times$ speedup compared to the Huggingface implementation through semantic-based sparse activation that identifies, fixes, and maintains stable neuron activation patterns at the sentence level. Laskaridis et al. introduced MELTing point \cite{laskaridis2024melting}, a mobile benchmarking suite designed to evaluate LLM performance, focusing on energy usage and memory footprints, across smartphones and Jetson platforms. TinyChatEngine \cite{tinyChatEngine} is also, an on-device LLM/VLM Inference Library that uses compression techniques to limit memory budgets while maintaining interactive response times on edge hardware. Furthermore, Nikoghosyan et al. showed that applying TensorRT to Transformer-based models on NVIDIA Jetson Xavier yields over 60\% latency reduction with negligible accuracy loss \cite{Nikoghosyan2023jetsonxaviertensorrt}.}

\Romina{In addition to language models, solutions target Super-Resolution (SR) and other vision-based generators. Chen et al. introduced TileSR \cite{chen2024tilesr}, which splits ultra-high-resolution images into tiles and selects the ones with the highest upscaling difficulty; these tiles are processed in parallel across multiple devices, reducing latency by up to 82\% and improving the image quality up to 10\% compared to other alternatives such as Supremo \cite{yi2022supremo} and MobiSR \cite{lee2019mobisr}. Wang et al.~\cite{wang2024intelligent} proposed ESHP, which combines a difficulty predictor with deep reinforcement learning to distribute SR tasks among CPUs, GPUs, and NPUs, speeding up SR processing without modifying the original architecture of the given SR model. Zhao et al. demonstrated a full-stack SR acceleration framework for embedded GPU devices, which outperformed standard TensorRT baselines in speed due to dictionary compression and operations optimization~\cite{zhao2021highperformance}.}

\Romina{FPGAs also provide a promising platform for runtime acceleration. Li et al. proposed a lookup-table (LUT)–based SR pipeline making sharper images while using much less energy without losing image quality \cite{li2024lut}. Other research has combined FFT-based processing with efficient multipliers \cite{chen2024fftwallace}, designed heterogeneous CNN-SNN architectures \cite{park2023resource}, or combined FPGA and GPU via PCIe to achieve real-time SR in microscopic imaging \cite{fang2022pcie}. For video-specific scenarios, Kim et al. employed pipeline and memory optimizations to reach 60~fps on 4K UHD content \cite{paris2019realtime}, while Sun et al. developed RNN compression techniques to manage temporal correlations \cite{zhang2022fpga_rnn}. %Larger multi-core systems also see performance gains: 
On larger multi-core systems Georgis et al. attained speedups over CPU-only baselines via parallelization \cite{abdelrahman2019acceleration}, and Liu et al. achieved real-time 4K SR on edge FPGAs through a DSP-enhanced caching scheme \cite{kumar2024highperformance}.}
\Romina{Finally, several system-level revisions help further reduce overhead. Fan et al.~\cite{liu2023covisu} leveraged codec-side data to skip redundant decoding in video SR, improved performance by up to 9.4$\times$. %compared to the traditional flow . 
Deformable 3D convolutional networks, essential in video tasks, were accelerated through tile decoupling and memory optimization by Zhang et al. \cite{zhou2022deformable3d}. Even resource-limited devices like the Raspberry Pi can support real-time SR: Osorno-Ortiz et al. integrated 2D-DWT with parallel interpolation to handle HD images in a short time \cite{liu2024dwt}.}

\section{Conclusion and Future Work}
\label{sec:5}
\Mozhgan{This work proposed a comprehensive survey regarding deploying Generative AI (GenAI) on edge devices. It presents a promising path toward reducing latency, enhancing data privacy, and enabling real-time capabilities in various applications. This survey has showcased the critical roles of software optimization, hardware specialization, and on-device inference frameworks in overcoming the resource constraints typical of embedded systems. Despite these advancements, significant challenges persist especially regarding model personalization, and security across distributed edge nodes. %Looking ahead, federated learning emerges as a compelling avenue for future work, allowing edge devices to collaboratively train or refine GenAI models without exposing raw data. 
By effectively addressing these challenges and combining these techniques with ongoing optimizations in model design and hardware acceleration, researchers and practitioners can pave the way for even more efficient, scalable, and privacy-preserving GenAI solutions at the edge.}

\begingroup
\tiny
\bibliography{main}

\begin{thebibliography}{138}
\providecommand{\natexlab}[1]{#1}

\bibitem[{{AI Squared}(2023)}]{ai2023dlite}
{AI Squared}. 2023.
\newblock {DLite V2}.
\newblock \url{https://huggingface.co/aisquared/dlite-v2-774m}.

\bibitem[{Ali et~al.(2024)}]{asmer2024energyv2}
Ali, A.~H.; et~al. 2024.
\newblock Energy-Aware FPGA Implementation of Spiking Neural Network with LIF Neurons.
\newblock \emph{arXiv preprint arXiv:2411.01628}.

\bibitem[{Beltagy et~al.(2020)}]{beltagy2020longformer}
Beltagy, I.; et~al. 2020.
\newblock {Longformer: The Long-Document Transformer}.
\newblock \emph{arXiv preprint arXiv:2004.05150}.

\bibitem[{Benmeziane and Maghraoui(2024)}]{10646428}
Benmeziane, H.; and Maghraoui, K.~E. 2024.
\newblock Are Large Language Models Good Neural Architecture Generators for Edge?
\newblock In \emph{2024 IEEE International Conference on Edge Computing and Communications (EDGE)}, 162--165.

\bibitem[{Bercovich et~al.(2024)}]{bercovich2024puzzle}
Bercovich, A.; et~al. 2024.
\newblock Puzzle: Distillation-Based NAS for Inference-Optimized LLMs.
\newblock \emph{arXiv preprint arXiv:2411.19146}.

\bibitem[{Black et~al.(2022)}]{black2022gpt}
Black, S.; et~al. 2022.
\newblock {GPT-NeoX-20B: An Open-Source Autoregressive Language Model}.
\newblock \emph{arXiv preprint arXiv:2204.06745}.

\bibitem[{Briq et~al.(2024)}]{briq2024datapruninggenerativediffusion}
Briq, R.; et~al. 2024.
\newblock Data Pruning in Generative Diffusion Models.
\newblock arXiv:2411.12523.

\bibitem[{Castells et~al.(2024)}]{Castells_2024_CVPR}
Castells, T.; et~al. 2024.
\newblock LD-Pruner: Efficient Pruning of Latent Diffusion Models using Task-Agnostic Insights.
\newblock In \emph{Proceedings of the IEEE/CVF Conference on Computer Vision and Pattern Recognition (CVPR) Workshops}, 821--830.

\bibitem[{Chen et~al.(2018)}]{chen2018regan}
Chen, F.; et~al. 2018.
\newblock {ReGAN: A pipelined ReRAM-based accelerator for generative adversarial networks}.
\newblock In \emph{2018 23rd Asia and South Pacific Design Automation Conference (ASP-DAC)}. IEEE.

\bibitem[{Chen et~al.(2024{\natexlab{a}})}]{chen2024tilesr}
Chen, N.; et~al. 2024{\natexlab{a}}.
\newblock {TileSR: Accelerate On-Device Super-Resolution with Parallel Offloading in Tile Granularity}.
\newblock In \emph{IEEE Annual International Conference on Computer Communications}.

\bibitem[{Chen et~al.(2024{\natexlab{b}})}]{chen2024channel}
Chen, Z.; et~al. 2024{\natexlab{b}}.
\newblock Channel-wise mixed-precision quantization for large language models.
\newblock \emph{arXiv preprint arXiv:2410.13056}.

\bibitem[{Chiang et~al.(2023)Chiang, Li, Lin, Sheng, Wu, Zhang, Zheng, Zhuang, Zhuang, Gonzalez, Stoica, and Xing}]{vicuna2023}
Chiang, W.-L.; Li, Z.; Lin, Z.; Sheng, Y.; Wu, Z.; Zhang, H.; Zheng, L.; Zhuang, S.; Zhuang, Y.; Gonzalez, J.~E.; Stoica, I.; and Xing, E.~P. 2023.
\newblock Vicuna: An Open-Source Chatbot Impressing GPT-4 with 90\%* ChatGPT Quality.

\bibitem[{Child et~al.(2019)}]{child2019sparse}
Child, R.; et~al. 2019.
\newblock {Generating Long Sequences with Sparse Transformers}.
\newblock \emph{arXiv preprint arXiv:1904.10509}.

\bibitem[{Choi et~al.(2023)}]{park2023resource}
Choi, J.; et~al. 2023.
\newblock {A Resource-Efficient Super-Resolution FPGA Processor with Heterogeneous CNN and SNN Core Architecture}.
\newblock \emph{IEEE Transactions on Computers}.

\bibitem[{Choromanski et~al.(2021)}]{choromanski2021performer}
Choromanski, K.; et~al. 2021.
\newblock {Rethinking Attention with Performers}.
\newblock In \emph{International Conference on Learning Representations (ICLR)}.

\bibitem[{Chung et~al.(2022)}]{won2022finetuned}
Chung, H.~W.; et~al. 2022.
\newblock {Scaling Instruction-Finetuned Language Models}.
\newblock \emph{arXiv preprint arXiv:2210.11416}.

\bibitem[{Computer(2023)}]{together2023redpajama}
Computer, T. 2023.
\newblock RedPajama: An Open Source Recipe to Reproduce LLaMA training dataset.

\bibitem[{Dao(2023)}]{dao2023flashattention2}
Dao, T. 2023.
\newblock {FlashAttention-2: Faster Attention with Better Parallelism and Work Partitioning}.
\newblock \emph{arXiv preprint arXiv:2307.08691}.

\bibitem[{Dao et~al.(2022)}]{dao2022flashattention}
Dao, T.; et~al. 2022.
\newblock {FlashAttention: Fast and Memory-Efficient Exact Attention with IO-Awareness}.
\newblock \emph{arXiv preprint arXiv:2205.14135}.

\bibitem[{DeepSeek-AI et~al.(2025)}]{deepseek2025}
DeepSeek-AI; et~al. 2025.
\newblock DeepSeek-R1: Incentivizing reasoning capability in LLMs via reinforcement learning.
\newblock \emph{arXiv preprint arXiv:2501.12948}.

\bibitem[{Dettmers et~al.(2023{\natexlab{a}})}]{dettmers2023qlora}
Dettmers, T.; et~al. 2023{\natexlab{a}}.
\newblock {QLoRA: Efficient Finetuning of Quantized LLMs}.
\newblock In \emph{Advances in Neural Information Processing Systems}, 10088--10115.

\bibitem[{Dettmers et~al.(2023{\natexlab{b}})}]{dettmers2023spqr}
Dettmers, T.; et~al. 2023{\natexlab{b}}.
\newblock {SpQR: A Sparse-Quantized Representation for Near-Lossless LLM Weight Compression}.
\newblock \emph{arXiv preprint arXiv:2306.03078}.
\newblock {Extended Preprint}.

\bibitem[{Dong et~al.(2024)}]{dong2024NLO}
Dong, P.; et~al. 2024.
\newblock Genetic Quantization-Aware Approximation for Non-Linear Operations in Transformers.
\newblock In \emph{Proceedings of the 61st ACM/IEEE Design Automation Conference (DAC)}.

\bibitem[{Egiazarian et~al.(2024)}]{egiazarian2024extreme}
Egiazarian, V.; et~al. 2024.
\newblock Extreme Compression of Large Language Models via Additive Quantization.
\newblock arXiv:2401.06118.

\bibitem[{Elsken et~al.(2019)}]{NASsurvey2019}
Elsken, T.; et~al. 2019.
\newblock Neural architecture search: A survey.
\newblock \emph{Journal of Machine Learning Research}.

\bibitem[{Fan et~al.(2023)}]{liu2023covisu}
Fan, H.; et~al. 2023.
\newblock {Co-ViSu: a Video Super-Resolution Accelerator Exploiting Codec Information Reuse}.
\newblock In \emph{International Conference on Field-Programmable Logic and Applications (FPL)}.

\bibitem[{Fang et~al.(2023)}]{fang2023structural}
Fang, G.; et~al. 2023.
\newblock Structural pruning for diffusion models.
\newblock In \emph{Advances in Neural Information Processing Systems}.

\bibitem[{Frantar and Alistarh(2023)}]{frantar-sparsegpt}
Frantar, E.; and Alistarh, D. 2023.
\newblock {SparseGPT}.
\newblock arXiv:2307.00026.

\bibitem[{Frantar et~al.(2022)}]{frantar2022gptq}
Frantar, E.; et~al. 2022.
\newblock {GPTQ: Accurate Post-Training Quantization for Generative Pre-trained Transformers}.
\newblock \emph{arXiv preprint arXiv:2210.17323}.

\bibitem[{Frantar et~al.(2023)}]{frantar2023optq}
Frantar, E.; et~al. 2023.
\newblock {OPTQ}: Accurate Quantization for Generative Pre-trained Transformers.
\newblock In \emph{The Eleventh International Conference on Learning Representations}.

\bibitem[{Fu et~al.(2024)}]{fu2024softact}
Fu, Y.; et~al. 2024.
\newblock SoftAct: A High-Precision Softmax Architecture for Transformers Supporting Nonlinear Functions.
\newblock \emph{IEEE Transactions on Circuits and Systems for Video Technology}, 34(9).

\bibitem[{Geng et~al.(2023{\natexlab{a}})}]{koala_blogpost_2023}
Geng, X.; et~al. 2023{\natexlab{a}}.
\newblock {Koala: A Dialogue Model for Academic Research.}

\bibitem[{Geng et~al.(2023{\natexlab{b}})}]{koala2023}
Geng, X.; et~al. 2023{\natexlab{b}}.
\newblock {Koala: A Dialogue Model for Academic Research}.
\newblock \url{https://bair.berkeley.edu/blog/2023/04/03/koala/}.

\bibitem[{Georgis et~al.(2019)}]{abdelrahman2019acceleration}
Georgis, G.; et~al. 2019.
\newblock {Acceleration techniques and evaluation on multi-core CPU, GPU and FPGA for image processing and super-resolution}.
\newblock \emph{Journal of Real-Time Image Processing}.

\bibitem[{Gu et~al.(2024)}]{gu2024minillm}
Gu, Y.; et~al. 2024.
\newblock Mini{LLM}: Knowledge Distillation of Large Language Models.
\newblock In \emph{The Twelfth International Conference on Learning Representations}.

\bibitem[{Gui et~al.(2022)}]{fang2022pcie}
Gui, D.; et~al. 2022.
\newblock {PCIe-based FPGA-GPU heterogeneous computation for real-time multi-emitter fitting in super-resolution localization microscopy}.
\newblock \emph{Biomedical Optics Express}.

\bibitem[{Guo et~al.(2023)Guo, Tang, Hu, Leng, Zhang, Yang, Liu, Guo, and Zhu}]{oliVe}
Guo, C.; Tang, J.; Hu, W.; Leng, J.; Zhang, C.; Yang, F.; Liu, Y.; Guo, M.; and Zhu, Y. 2023.
\newblock Olive: Accelerating large language models via hardware-friendly outlier-victim pair quantization.
\newblock In \emph{Proceedings of the 50th Annual International Symposium on Computer Architecture}, 1--15.

\bibitem[{Guo et~al.(2022)}]{guo2022ant}
Guo, C.; et~al. 2022.
\newblock Ant: Exploiting adaptive numerical data type for low-bit deep neural network quantization.
\newblock In \emph{2022 55th IEEE/ACM International Symposium on Microarchitecture (MICRO)}, 1414--1433. IEEE.

\bibitem[{Guo et~al.(2024{\natexlab{a}})}]{guo2024survey}
Guo, C.; et~al. 2024{\natexlab{a}}.
\newblock A Survey: Collaborative Hardware and Software Design in the Era of Large Language Models.
\newblock \emph{arXiv preprint arXiv:2410.07265}.

\bibitem[{Guo et~al.(2024{\natexlab{b}})}]{guo2024hcaeds}
Guo, R.; et~al. 2024{\natexlab{b}}.
\newblock 20.2 A 28nm 74.34TFLOPS/W BF16 Heterogenous CIM-Based Accelerator Exploiting Denoising-Similarity for Diffusion Models.
\newblock In \emph{2024 IEEE International Solid-State Circuits Conference (ISSCC)}, volume~67, 362--364.

\bibitem[{He et~al.(2023)}]{ptqd}
He, Y.; et~al. 2023.
\newblock PTQD: Accurate Post-Training Quantization for Diffusion Models.
\newblock In Oh, A.; Naumann, T.; Globerson, A.; Saenko, K.; Hardt, M.; and Levine, S., eds., \emph{Advances in Neural Information Processing Systems}, volume~36, 13237--13249. Curran Associates, Inc.

\bibitem[{Heo et~al.(2025)}]{heo2025exion}
Heo, J.; et~al. 2025.
\newblock EXION: Exploiting Inter-and Intra-Iteration Output Sparsity for Diffusion Models.
\newblock \emph{arXiv preprint arXiv:2501.05680}.

\bibitem[{Howard et~al.(2017)}]{howard2017mobilenets}
Howard, A.~G.; et~al. 2017.
\newblock Mobilenets: Efficient convolutional neural networks for mobile vision applications.
\newblock \emph{arXiv preprint arXiv:1704.04861}.

\bibitem[{Hu et~al.(2021)}]{hulora2021}
Hu, E.; et~al. 2021.
\newblock {LoRA: Low-Rank Adaptation of Large Language Models}.
\newblock \emph{arXiv preprint arXiv:2106.09685}.

\bibitem[{Huang et~al.(2024)}]{Huang_2024_CVPR}
Huang, Y.; et~al. 2024.
\newblock {TFMQ-DM: Temporal Feature Maintenance Quantization for Diffusion Models}.
\newblock In \emph{{Proceedings of the IEEE/CVF Conference on Computer Vision and Pattern Recognition (CVPR)}}, 7362--7371.

\bibitem[{Humes et~al.(2023)}]{humes2023squeezedv2}
Humes, E.; et~al. 2023.
\newblock Squeezed Edge YOLO: Onboard Object Detection on Edge Devices.
\newblock \emph{arXiv preprint arXiv:2312.11716}.

\bibitem[{Jiao et~al.(2020)}]{jiao2019tinybert}
Jiao, X.; et~al. 2020.
\newblock {TinyBERT: Distilling BERT for Natural Language Understanding}.
\newblock \emph{arXiv preprint arXiv:1909.10351}.

\bibitem[{Kallakuri et~al.(2024)}]{uttej2024resourcev2}
Kallakuri, U.; et~al. 2024.
\newblock Resource-Aware Saliency-Guided Differentiable Pruning for Deep Neural Networks.
\newblock In \emph{Proceedings of the Great Lakes Symposium on VLSI 2024}, 694--699.

\bibitem[{Kang et~al.(2021)}]{kang2021ganpu}
Kang, S.; et~al. 2021.
\newblock GANPU: An Energy-Efficient Multi-DNN Training Processor for GANs With Speculative Dual-Sparsity Exploitation.
\newblock \emph{IEEE Journal of Solid-State Circuits}, 56(9): 2845--2857.

\bibitem[{Khiabani et~al.(2025)}]{khiabani2025optimizing}
Khiabani, Y.~S.; et~al. 2025.
\newblock Optimizing Small Language Models for In-Vehicle Function-Calling.
\newblock \emph{arXiv preprint arXiv:2501.02342}.

\bibitem[{Kim et~al.(2020)}]{kim2021gan}
Kim, S.; et~al. 2020.
\newblock An Energy-Efficient GAN Accelerator with On-chip Training for Domain Specific Optimization.
\newblock In \emph{2020 IEEE Asian Solid-State Circuits Conference (A-SSCC)}, 1--4.

\bibitem[{Kim et~al.(2024)}]{kim2024squeezellm}
Kim, S.; et~al. 2024.
\newblock {SqueezeLLM: Dense-and-Sparse Quantization}.
\newblock In \emph{Proceedings of the 41st International Conference on Machine Learning (ICML)}.

\bibitem[{Kim et~al.(2019)}]{paris2019realtime}
Kim, Y.; et~al. 2019.
\newblock {A Real-Time Convolutional Neural Network for Super-Resolution on FPGA With Applications to 4K UHD 60 fps Video Services}.
\newblock \emph{IEEE Transactions on Circuits and Systems for Video Technology}.

\bibitem[{Kitaev et~al.(2020)}]{kitaev2020reformer}
Kitaev, N.; et~al. 2020.
\newblock {Reformer: The Efficient Transformer}.
\newblock In \emph{International Conference on Learning Representations (ICLR)}.

\bibitem[{Kong et~al.(2024)}]{kong2024camb}
Kong, W.; et~al. 2024.
\newblock Cambricon-D: Full-Network Differential Acceleration for Diffusion Models.
\newblock In \emph{2024 ACM/IEEE 51st Annual International Symposium on Computer Architecture (ISCA)}.

\bibitem[{Lan et~al.(2020)}]{lan2019albert}
Lan, Z.; et~al. 2020.
\newblock {ALBERT: A Lite BERT for Self-supervised Learning of Language Representations}.
\newblock In \emph{International Conference on Learning Representations (ICLR)}.

\bibitem[{Laskaridis et~al.(2024)}]{laskaridis2024melting}
Laskaridis, S.; et~al. 2024.
\newblock {MELTing point: Mobile Evaluation of Language Transformers}.
\newblock \emph{arXiv preprint arXiv:2403.12844}.

\bibitem[{Lee et~al.(2023)}]{lee2023rlaif}
Lee, H.; et~al. 2023.
\newblock RLAIF: Scaling Reinforcement Learning from Human Feedback with AI Feedback.
\newblock arXiv:2309.00267.

\bibitem[{Lee et~al.(2019)}]{lee2019mobisr}
Lee, R.; et~al. 2019.
\newblock {MobiSR: Efficient on-device super-resolution through heterogeneous mobile processors}.
\newblock In \emph{Proceedings of the 25th Annual International Conference on Mobile Computing and Networking (MobiCom)}.

\bibitem[{Lefaudeux et~al.(2022)}]{xformer}
Lefaudeux, B.; et~al. 2022.
\newblock xFormers: A modular and hackable Transformer modelling library.
\newblock \url{https://github.com/facebookresearch/xformers}.

\bibitem[{Li et~al.(2024{\natexlab{a}})}]{li2024specpim}
Li, C.; et~al. 2024{\natexlab{a}}.
\newblock {SpecPIM: Accelerating Speculative Inference on PIM-Enabled System via Architecture-Dataflow Co-Exploration}.
\newblock In \emph{{Proceedings of the 29th ACM International Conference on Architectural Support for Programming Languages and Operating Systems, Volume 3}}.

\bibitem[{Li et~al.(2024{\natexlab{b}})}]{li2024lut}
Li, H.; et~al. 2024{\natexlab{b}}.
\newblock {An Energy-Efficient Look-up Table Framework for Super Resolution on FPGA}.
\newblock \emph{IEEE Transactions on Circuits and Systems for Video Technology}.

\bibitem[{Li et~al.(2024{\natexlab{c}})}]{li2024asadi}
Li, H.; et~al. 2024{\natexlab{c}}.
\newblock {ASADI: Accelerating Sparse Attention Using Diagonal-based In-Situ Computing}.
\newblock In \emph{{2024 IEEE International Symposium on High-Performance Computer Architecture (HPCA)}}. IEEE.

\bibitem[{Li et~al.(2023{\natexlab{a}})}]{q-diffusion}
Li, X.; et~al. 2023{\natexlab{a}}.
\newblock Q-Diffusion: Quantizing Diffusion Models.
\newblock In \emph{Proceedings of the IEEE/CVF International Conference on Computer Vision (ICCV)}, 17535--17545.

\bibitem[{Li et~al.(2023{\natexlab{b}})}]{li2023sparse}
Li, Y.; et~al. 2023{\natexlab{b}}.
\newblock {E-sparse: Boosting the large language model inference through entropy-based n: M sparsity.}
\newblock arXiv:2310.12843.

\bibitem[{Li et~al.(2023{\natexlab{c}})}]{q-dm}
Li, Y.; et~al. 2023{\natexlab{c}}.
\newblock {Q-DM: An Efficient Low-bit Quantized Diffusion Model}.
\newblock In \emph{{Advances in Neural Information Processing Systems}}, 76680--76691.

\bibitem[{Liang et~al.(2024)}]{liang2024dynamic}
Liang, Z.; et~al. 2024.
\newblock Dynamic Self-adaptive Multiscale Distillation from Pre-trained Multimodal Large Model for Efficient Cross-modal Representation Learning.
\newblock \emph{arXiv preprint arXiv:2404.10838}.

\bibitem[{Lin et~al.(2024{\natexlab{a}})}]{lin2023awq}
Lin, J.; et~al. 2024{\natexlab{a}}.
\newblock AWQ: Activation-aware Weight Quantization for On-Device LLM Compression and Acceleration.
\newblock In Gibbons, P.; Pekhimenko, G.; and Sa, C.~D., eds., \emph{Proceedings of Machine Learning and Systems}, volume~6, 87--100.

\bibitem[{Lin et~al.(2024{\natexlab{b}})}]{lin2024speechprune}
Lin, Y.; et~al. 2024{\natexlab{b}}.
\newblock SpeechPrune: Context-aware Token Pruning for Speech Information Retrieval.
\newblock \emph{arXiv preprint arXiv:2412.12009}.

\bibitem[{Liu et~al.(2024{\natexlab{a}})}]{liu-etal-2024-evolving}
Liu, C.; et~al. 2024{\natexlab{a}}.
\newblock Evolving Knowledge Distillation with Large Language Models and Active Learning.
\newblock In \emph{Proceedings of the 2024 Joint International Conference on Computational Linguistics, Language Resources and Evaluation (LREC-COLING 2024)}, 6717--6731. ELRA and ICCL.

\bibitem[{Liu et~al.(2024{\natexlab{b}})}]{kumar2024highperformance}
Liu, H.; et~al. 2024{\natexlab{b}}.
\newblock {A High-Performance Accelerator for Real-Time Super-Resolution on Edge FPGAs}.
\newblock \emph{ACM Transactions on Design Automation of Electronic Systems}.

\bibitem[{Liu et~al.(2024{\natexlab{c}})}]{liu2024mobilellm}
Liu, Z.; et~al. 2024{\natexlab{c}}.
\newblock Mobilellm: Optimizing sub-billion parameter language models for on-device use cases.
\newblock \emph{arXiv preprint arXiv:2402.14905}.

\bibitem[{Lu et~al.(2022)}]{NEURIPS2022_260a14ac}
Lu, C.; et~al. 2022.
\newblock DPM-Solver: A Fast ODE Solver for Diffusion Probabilistic Model Sampling in Around 10 Steps.
\newblock In Koyejo, S.; Mohamed, S.; Agarwal, A.; Belgrave, D.; Cho, K.; and Oh, A., eds., \emph{Advances in Neural Information Processing Systems}, 5775--5787. Curran Associates, Inc.

\bibitem[{Luhman and Luhman(2021)}]{luhman2021knowledge}
Luhman, E.; and Luhman, T. 2021.
\newblock Knowledge distillation in iterative generative models for improved sampling speed.
\newblock \emph{arXiv preprint arXiv:2101.02388}.

\bibitem[{Ma et~al.(2023)}]{ma2023llmpruner}
Ma, X.; et~al. 2023.
\newblock LLM-Pruner: On the Structural Pruning of Large Language Models.
\newblock In \emph{Advances in Neural Information Processing Systems}.

\bibitem[{Malathi et~al.(2024)}]{chen2024fftwallace}
Malathi, L.; et~al. 2024.
\newblock {FPGA design of FFT-based intelligent accelerator with optimized Wallace tree multiplier for image super resolution and quality enhancement}.
\newblock \emph{Biomedical Signal Processing and Control}.

\bibitem[{Manjunath et~al.(2023)}]{reprohrl2023-tejaswiniv2}
Manjunath, T.; et~al. 2023.
\newblock Reprohrl: Towards multi-goal navigation in the real world using hierarchical agents. On 37th AAAI Conference on Artificial Intelligence.
\newblock In \emph{The 1st Reinforcement Learning Ready for Production workshop}.

\bibitem[{Mao et~al.(2017{\natexlab{a}})}]{mao2}
Mao, J.; et~al. 2017{\natexlab{a}}.
\newblock {MeDNN: A distributed mobile system with enhanced partition and deployment for large-scale DNNs}.
\newblock In \emph{{2017 IEEE/ACM International Conference on Computer-Aided Design (ICCAD)}}, 751--756. IEEE.

\bibitem[{Mao et~al.(2017{\natexlab{b}})}]{mao1}
Mao, J.; et~al. 2017{\natexlab{b}}.
\newblock MoDNN: Local distributed mobile computing system for Deep Neural Network.
\newblock In \emph{Design, Automation \& Test in Europe Conference \& Exhibition (DATE), 2017}, 1396--1401.

\bibitem[{Meng et~al.(2023)}]{meng2023distillation}
Meng, C.; et~al. 2023.
\newblock On distillation of guided diffusion models.
\newblock In \emph{Proceedings of the IEEE/CVF Conference on Computer Vision and Pattern Recognition}, 14297--14306.

\bibitem[{MIT-HAN-Lab(2024)}]{tinyChatEngine}
MIT-HAN-Lab. 2024.
\newblock TinyChatEngine: On-Device LLM Inference Library.
\newblock \url{https://github.com/mit-han-lab/TinyChatEngine}.

\bibitem[{{MosaicML NLP Team}(2023)}]{MosaicML2023Introducing}
{MosaicML NLP Team}. 2023.
\newblock {Introducing MPT-7B: A new standard for open-source, commercially usable LLMs}.
\newblock \url{https://www.mosaicml.com/blog/mpt-7b}.

\bibitem[{Navardi et~al.(2023)}]{iscas2023-mozhganv2}
Navardi, M.; et~al. 2023.
\newblock MLAE2: Metareasoning for latency-aware energy-efficient autonomous nano-drones.
\newblock In \emph{2023 IEEE International Symposium on Circuits and Systems (ISCAS)}, 1--5. IEEE.

\bibitem[{Navardi et~al.(2024)}]{mozhgan2024metatinymlv2}
Navardi, M.; et~al. 2024.
\newblock MetaTinyML: End-to-End Metareasoning Framework for TinyML Platforms.
\newblock \emph{IEEE Embedded Systems Letters}, 16(4): 393--396.

\bibitem[{Nezami et~al.(2024)}]{nezami2024generative}
Nezami, Z.; et~al. 2024.
\newblock Generative AI on the Edge: Architecture and Performance Evaluation.
\newblock \emph{arXiv preprint arXiv:2411.17712}.

\bibitem[{Nikoghosyan et~al.(2023)}]{Nikoghosyan2023jetsonxaviertensorrt}
Nikoghosyan, K.~H.; et~al. 2023.
\newblock {Acceleration of Transformer Architectures on Jetson Xavier using TensorRT}.
\newblock In \emph{Proceedings of Innovative Polytechnic}.

\bibitem[{{NVIDIA Corporation}(2025)}]{tensorrtllm}
{NVIDIA Corporation}. 2025.
\newblock {TensorRT-LLM}.
\newblock \url{https://github.com/NVIDIA/TensorRT-LLM}.

\bibitem[{{OpenAI}(2024)}]{openai2024gpt4technicalreport}
{OpenAI}. 2024.
\newblock {GPT-4 Technical Report}.
\newblock arXiv:2303.08774.

\bibitem[{Osorno-Ortiz et~al.(2024)}]{liu2024dwt}
Osorno-Ortiz, R.~J.; et~al. 2024.
\newblock {Implementation of the image super-resolution DWT based algorithm on Raspberry Pi platform for real-time applications}.
\newblock In \emph{Proceedings of SPIE}.

\bibitem[{Park et~al.(2024)}]{park2024attacc}
Park, J.; et~al. 2024.
\newblock {AttAcc! Unleashing the Power of PIM for Batched Transformer-based Generative Model Inference}.
\newblock In \emph{{Proceedings of the 29th ACM International Conference on Architectural Support for Programming Languages and Operating Systems, Volume 2}}.

\bibitem[{Poole et~al.(2023)}]{poole2023dreamfusion}
Poole, B.; et~al. 2023.
\newblock DreamFusion: Text-to-3D using 2D Diffusion.
\newblock In \emph{The Eleventh International Conference on Learning Representations}.

\bibitem[{Pourmehrani et~al.(2024)}]{pourmehrani2024fat}
Pourmehrani, H.; et~al. 2024.
\newblock FAT-RABBIT: Fault-Aware Training towards Robustness AgainstBit-flip Based Attacks in Deep Neural Networks.
\newblock In \emph{2024 IEEE International Test Conference (ITC)}, 106--110. IEEE.

\bibitem[{Qin et~al.(2024{\natexlab{a}})}]{FLNAS2024}
Qin, R.; et~al. 2024{\natexlab{a}}.
\newblock FL-NAS: Towards Fairness of NAS for Resource Constrained Devices via Large Language Models.
\newblock In \emph{Proceedings of the 29th Asia and South Pacific Design Automation Conference}, ASPDAC '24, 429–434. IEEE Press.
\newblock ISBN 9798350393545.

\bibitem[{Qin et~al.(2023)}]{qin2023FACT}
Qin, Y.; et~al. 2023.
\newblock FACT: FFN-Attention Co-optimized Transformer Architecture with Eager Correlation Prediction.
\newblock In \emph{Proceedings of the 50th Annual International Symposium on Computer Architecture}, ISCA '23. New York, NY, USA: Association for Computing Machinery.
\newblock ISBN 9798400700958.

\bibitem[{Qin et~al.(2024{\natexlab{b}})}]{qin2024dmpu}
Qin, Y.; et~al. 2024{\natexlab{b}}.
\newblock A 52.01 TFLOPS/W Diffusion Model Processor with Inter-Time-Step Convolution-Attention-Redundancy Elimination and Bipolar Floating-Point Multiplication.
\newblock In \emph{2024 IEEE Symposium on VLSI Technology and Circuits (VLSI Technology and Circuits)}, 1--2.

\bibitem[{Qin et~al.(2024{\natexlab{c}})}]{qin2024mecla}
Qin, Y.; et~al. 2024{\natexlab{c}}.
\newblock MECLA: Memory-Compute-Efficient LLM Accelerator with Scaling Sub-matrix Partition.
\newblock In \emph{2024 ACM/IEEE 51st Annual International Symposium on Computer Architecture (ISCA)}, 1032--1047.

\bibitem[{Salimans and Ho(2022)}]{salimans2022progressive}
Salimans, T.; and Ho, J. 2022.
\newblock Progressive Distillation for Fast Sampling of Diffusion Models.
\newblock In \emph{International Conference on Learning Representations}.

\bibitem[{Sanh et~al.(2019)}]{sanh2019distilbert}
Sanh, V.; et~al. 2019.
\newblock {DistilBERT, a Distilled Version of BERT: Smaller, Faster, Cheaper and Lighter}.
\newblock \emph{arXiv preprint arXiv:1910.01108}.

\bibitem[{Shah et~al.(2024)}]{dao2024flashattention3}
Shah, J.; et~al. 2024.
\newblock {FlashAttention-3: Fast and Accurate Attention with Asynchrony and Low-precision}.
\newblock \emph{arXiv preprint arXiv:2407.08608}.

\bibitem[{Shen et~al.(2024)}]{shen2024edgeqat}
Shen, X.; et~al. 2024.
\newblock EdgeQAT: Entropy and Distribution Guided Quantization-Aware Training for the Acceleration of Lightweight LLMs on the Edge.
\newblock \emph{arXiv preprint arXiv:2402.10787}.

\bibitem[{{Singer, Philipp others}(2024)}]{h2odanube2024}
{Singer, Philipp others}. 2024.
\newblock {H2O-Danube-1.8B Technical Report}.
\newblock \emph{arXiv preprint arXiv:2401.16818}.

\bibitem[{So et~al.(2023)}]{NEURIPS2023_983591c3}
So, J.; et~al. 2023.
\newblock Temporal Dynamic Quantization for Diffusion Models.
\newblock In Oh, A.; Naumann, T.; Globerson, A.; Saenko, K.; Hardt, M.; and Levine, S., eds., \emph{Advances in Neural Information Processing Systems}, volume~36, 48686--48698. Curran Associates, Inc.

\bibitem[{Song et~al.(2023)}]{pmlr-v202-song23a}
Song, Y.; et~al. 2023.
\newblock {Consistency Models}.
\newblock In \emph{{Proceedings of the 40th International Conference on Machine Learning}}, 32211--32252.

\bibitem[{Stevens et~al.(2021)}]{ste2021softermax}
Stevens, J.~R.; et~al. 2021.
\newblock Softermax: Hardware/Software Co-Design of an Efficient Softmax for Transformers.
\newblock In \emph{2021 58th ACM/IEEE Design Automation Conference (DAC)}, 469--474.

\bibitem[{Su et~al.(2024)}]{su2024diffusion}
Su, D.; et~al. 2024.
\newblock Generative Dataset Distillation Based on Diffusion Model.
\newblock In \emph{Proceedings of the European Conference on Computer Vision (ECCV), Workshop}.

\bibitem[{Sun et~al.(2022)}]{zhang2022fpga_rnn}
Sun, K.; et~al. 2022.
\newblock {An FPGA-Based Residual Recurrent Neural Network for Real-Time Video Super-Resolution}.
\newblock \emph{IEEE Transactions on Circuits and Systems for Video Technology}.

\bibitem[{Sun et~al.(2020)}]{sun2020mobilebert}
Sun, Z.; et~al. 2020.
\newblock {MobileBERT: a Compact Task-Agnostic BERT for Resource-Limited Devices}.
\newblock \emph{arXiv preprint arXiv:2004.02984}.

\bibitem[{Tambe et~al.(2023)}]{tambe2023STP}
Tambe, T.; et~al. 2023.
\newblock {22.9 A 12nm 18.1 TFLOPs/W sparse transformer processor with entropy-based early exit, mixed-precision predication and fine-grained power management}.
\newblock In \emph{{2023 IEEE International Solid-State Circuits Conference (ISSCC)}}, 370--372. IEEE.

\bibitem[{Taori et~al.(2023)}]{alpaca2023}
Taori, R.; et~al. 2023.
\newblock {Alpaca: A Strong, Replicable Instruction-Following Model}.
\newblock \url{https://crfm.stanford.edu/2023/03/13/alpaca.html}.

\bibitem[{Touvron et~al.(2023{\natexlab{a}})}]{touvron2023llama2}
Touvron, H.; et~al. 2023{\natexlab{a}}.
\newblock {Llama 2: Open Foundation and Fine-Tuned Chat Models}.
\newblock \emph{arXiv preprint arXiv:2307.09288}.

\bibitem[{Touvron et~al.(2023{\natexlab{b}})}]{touvron2023llamaopenefficientfoundation}
Touvron, H.; et~al. 2023{\natexlab{b}}.
\newblock {LLaMA: Open and Efficient Foundation Language Models}.
\newblock In \emph{{arXiv preprint arXiv:2302.13971}}.

\bibitem[{Wan et~al.(2025)}]{wan2025pruningsparsediffusionmodels}
Wan, B.; et~al. 2025.
\newblock {Pruning for Sparse Diffusion Models based on Gradient Flow.}
\newblock arXiv:2501.01101.

\bibitem[{Wang et~al.(2023)}]{wang2023bitnet}
Wang, H.; et~al. 2023.
\newblock {Bitnet: Scaling 1-bit transformers for large language models.}
\newblock arXiv:2305.10403.

\bibitem[{Wang et~al.(2024{\natexlab{a}})}]{wang2024intelligent}
Wang, Q.; et~al. 2024{\natexlab{a}}.
\newblock {An Intelligent Co-Scheduling Framework for Efficient Super-Resolution on Edge Platforms With Heterogeneous Processors}.
\newblock \emph{IEEE Internet of Things Journal}.

\bibitem[{Wang et~al.(2024{\natexlab{b}})}]{wang2024coreinferacceleratinglargelanguage}
Wang, Q.; et~al. 2024{\natexlab{b}}.
\newblock {CoreInfer: Accelerating Large Language Model Inference with Semantics-Inspired Adaptive Sparse Activation}.
\newblock arXiv:2410.18311.

\bibitem[{Wang et~al.(2020{\natexlab{a}})}]{wang2020linformer}
Wang, S.; et~al. 2020{\natexlab{a}}.
\newblock {Linformer: Self-Attention with Linear Complexity}.
\newblock \emph{arXiv preprint arXiv:2006.04768}.

\bibitem[{Wang et~al.(2020{\natexlab{b}})}]{wang2020minilmdeepselfattentiondistillation}
Wang, W.; et~al. 2020{\natexlab{b}}.
\newblock MiniLM: Deep Self-Attention Distillation for Task-Agnostic Compression of Pre-Trained Transformers.

\bibitem[{Wang et~al.(2021)}]{wang2021minilmv2multiheadselfattentionrelation}
Wang, W.; et~al. 2021.
\newblock MiniLMv2: Multi-Head Self-Attention Relation Distillation for Compressing Pretrained Transformers.

\bibitem[{Wang et~al.(2024{\natexlab{c}})}]{wang2024dtrans}
Wang, X.; et~al. 2024{\natexlab{c}}.
\newblock {DTrans: A Dataflow-transformation FPGA Accelerator with Nonlinear-operators fusion aiming for the Generative Model}.
\newblock In \emph{{2024 34th International Conference on Field-Programmable Logic and Applications (FPL)}}. IEEE.

\bibitem[{Wu et~al.(2024)}]{wu-etal-2024-lamini}
Wu, M.; et~al. 2024.
\newblock {Lamini: Large Language Model Approaches for Generating Assistive Programs}.
\newblock In \emph{{Proceedings of the 18th Conference of the European Chapter of the Association for Computational Linguistics (Volume 1: Long Papers)}}, 944--964.

\bibitem[{Xia et~al.(2023)}]{xia2023flashllm}
Xia, H.; et~al. 2023.
\newblock {Flash-LLM: Enabling Cost-Effective and Highly-Efficient Large Generative Model Inference with Unstructured Sparsity.}
\newblock arXiv:2308.04528.

\bibitem[{Xiao et~al.(2023)}]{xiao2023smoothquant}
Xiao, G.; et~al. 2023.
\newblock {S}mooth{Q}uant: Accurate and Efficient Post-Training Quantization for Large Language Models.
\newblock In \emph{Proceedings of the 40th International Conference on Machine Learning}.

\bibitem[{Xu et~al.(2024)}]{xu2024onebit}
Xu, Y.; et~al. 2024.
\newblock OneBit: Towards Extremely Low-bit Large Language Models.
\newblock \emph{arXiv preprint arXiv:2402.11295}.

\bibitem[{Yan et~al.(2019)}]{yan2019cim_non}
Yan, B.; et~al. 2019.
\newblock {RRAM-based spiking nonvolatile computing-in-memory processing engine with precision-configurable in situ nonlinear activation}.
\newblock In \emph{{2019 Symposium on VLSI Technology}}, C258--C259. IEEE.

\bibitem[{Yan et~al.(2024)}]{yan2024hybrid}
Yan, C.; et~al. 2024.
\newblock {Hybrid SD: Edge-Cloud Collaborative Inference for Stable Diffusion Models.}
\newblock arXiv:2410.02453.

\bibitem[{Yang et~al.(2024)}]{yang2024sda}
Yang, G.; et~al. 2024.
\newblock {SDA: Low-Bit Stable Diffusion Acceleration on Edge FPGAs}.
\newblock In \emph{{2024 34th International Conference on Field-Programmable Logic and Applications (FPL)}}. IEEE.

\bibitem[{Yao et~al.(2024)}]{yao2024timestep}
Yao, Y.; et~al. 2024.
\newblock Timestep-Aware Correction for Quantized Diffusion Models.
\newblock In \emph{European Conference on Computer Vision (ECCV)}.

\bibitem[{Yi et~al.(2022)}]{yi2022supremo}
Yi, J.; et~al. 2022.
\newblock {Supremo: Cloud-assisted low-latency super-resolution in mobile devices}.
\newblock \emph{IEEE Transactions on Mobile Computing}.

\bibitem[{Yi et~al.(2023)}]{yi2023edgemoe}
Yi, R.; et~al. 2023.
\newblock {EdgeMoE: Fast On-Device Inference of MoE-Based Large Language Models}.
\newblock \emph{arXiv preprint arXiv:2308.14352}.

\bibitem[{Yoo et~al.(2024)}]{yoo2024eeda}
Yoo, S.; et~al. 2024.
\newblock A 28nm 4.96 TOPS/W End-to-End Diffusion Accelerator with Reconfigurable Hyper-Precision and Unified Non-Matrix Processing Engine.
\newblock In \emph{2024 IEEE European Solid-State Electronics Research Conference (ESSERC)}, 253--256.

\bibitem[{Zadeh et~al.(2020)}]{zad2020gobo}
Zadeh, A.~H.; et~al. 2020.
\newblock GOBO: Quantizing Attention-Based NLP Models for Low Latency and Energy Efficient Inference.
\newblock In \emph{2020 53rd Annual IEEE/ACM International Symposium on Microarchitecture (MICRO)}, 811--824.

\bibitem[{Zaheer et~al.(2020)}]{zaheer2020bigbird}
Zaheer, M.; et~al. 2020.
\newblock {Big Bird: Transformers for Longer Sequences}.
\newblock In \emph{Advances in Neural Information Processing Systems}.

\bibitem[{Zhang et~al.(2024)}]{tinyllama2024}
Zhang, P.; et~al. 2024.
\newblock {TinyLlama: An Open-Source Small Language Model}.
\newblock \emph{arXiv preprint arXiv:2401.02385}.

\bibitem[{Zhang et~al.(2022)}]{zhou2022deformable3d}
Zhang, S.; et~al. 2022.
\newblock {An Efficient Accelerator of Deformable 3D Convolutional Network for Video Super-Resolution}.
\newblock \emph{IEEE Transactions on Multimedia}.

\bibitem[{Zhao et~al.(2021)}]{zhao2021highperformance}
Zhao, W.; et~al. 2021.
\newblock {A High-Performance Accelerator for Super-Resolution Processing on Embedded GPU}.
\newblock \emph{arXiv preprint arXiv:2303.08999}.

\bibitem[{Zhu et~al.(2024)}]{zhu2024dipgodiffusionprunerfewstep}
Zhu, H.; et~al. 2024.
\newblock DiP-GO: A Diffusion Pruner via Few-step Gradient Optimization.
\newblock \emph{arXiv preprint arXiv:2410.16942}.

\bibitem[{Zoph(2016)}]{NASfirst2016}
Zoph, B. 2016.
\newblock Neural architecture search with reinforcement learning.
\newblock \emph{arXiv preprint arXiv:1611.01578}.

\bibitem[{Zoph et~al.(2018)}]{NASextended2018}
Zoph, B.; et~al. 2018.
\newblock Learning transferable architectures for scalable image recognition.
\newblock In \emph{Proceedings of the IEEE conference on computer vision and pattern recognition}, 8697--8710.

\end{thebibliography}
\endgroup
\end{document}